\documentclass[published]{JHEP3} 

\JHEP{00(0000)000}

\JHEPspecialurl{http://jhep.sissa.it/JOURNAL/JHEP3.tar.gz}

\usepackage{epsfig,multicol,bbm}

\newcommand{\bean}{\begin{eqnarray}}

\newcommand{\eean}{\end{eqnarray}}

\newcommand{\eq}[1]{Eq. (\ref{#1})}
\newcommand{\meq}[1]{(\ref{#1})}

\newcommand{\abp}{A_{B'}}

\newcommand\fverb{\setbox\fverbbox=\hbox\bgroup\verb}
\newcommand\fverbdo{\egroup\medskip\noindent%
            \fbox{\unhbox\fverbbox}\ }
\newcommand\fverbit{\egroup\item[\fbox{\unhbox\fverbbox}]}
\newbox\fverbbox

\title{\bf Proof of the entropy bound on dynamical horizons}

\author{ Sijie Gao, \\
\\Department of Physics, \\Beijing Normal University,\\
Beijing 100875, China \\ Email: \email{sijie@bnu.edu.cn}}
\author{  Xiaoning Wu   \\
 Institute of Applied Mathematics, \\Academy of Mathematics and
System Science, \\Chinese Academy of Sciences, Beijing,
100080,China. \\Email: \email{wuxn@amss.ac.cn}}

\received{}       
\accepted{}       

\abstract
 {The entropy bound conjecture concerning black hole
dynamical horizons is proved. The conjecture states, if a dynamical
horizon, $D_H$,  is bounded by two surfaces with areas of $A_B$ and
$\abp$ ($\abp>A_B$), then the entropy, $S_D$, that crosses $D_H$
must satisfy $S_D\leq \frac{1}{4}(\abp-A_B)$. We show that this
conjecture is implied by the generalized Bousso bound. Consequently,
the generalized second law holds for dynamical horizons. Finally, we
show that the lightlike bousso bound and its spacelike counterpart
can be unified as one bound.}


\begin{document}

\section{Introduction}
Bekenstein \cite{beken81} has conjectured that the entropy $S$ and
energy $E$ of any thermodynamic system must obey
 \bean
 S \leq2\pi E R \,, \label{bbound}
\eean
where $R$ is defined as the circumferential radius. This bound
is universal in the sense that it is supposed to hold in any matter
system. The Bekenstein bound has been confirmed in wide classes of
systems. However, as pointed by Bekenstein, the bound is valid for
 systems with finite size and limited self-gravity.
Counterexamples can be easily found in systems undergoing
gravitational collapse \cite{bousso}. Another entropy bound is
related to the holographic principle, which says that the entropy
in a spherical volume satisfies
\bean S\leq \frac{A}{4} \,,  \label{areabound}
\eean where $A$ is
the area of the system. It was shown that this bound
 is violated for sufficiently large volumes
\cite{susskind}. Note that the entropy mentioned above is
contained in a spacelike region. To find an entropy bound that has
wider applications, Bousso investigated the entropy crossing a
lightlike hypersurface and proposed a covariant entropy bound
conjecture. Consider a spacelike 2-surface $B$ with area  $A_B$ in
a spacetime satisfying the dominant energy condition. A null
hypersurface $L$ is generated by null geodesics, which starts at
$B$ and is orthogonal to $B$. Let $k^a$ be the tangent vector
field of the geodesics and $\theta$ the expansion associated with
$k^a$. Suppose that $\theta$ is nonpositive everywhere on $L$ and
not terminated until a caustic is reached. Then the entropy,
$S_L$, that crosses $L$ satisfies 
\bean S_L\leq \frac{1}{4}A_B \,. \label{boussobound} \eean
\eq{boussobound} is the covariant entropy bound proposed by Bousso.
This bound has passed the tests in various cases \cite{bousso,
wald3, gao,gaolocal}. A generalized Bousso bound allows $L$ to be
terminated at any spacelike 2-surface $B'$ with area $A_{B'}$ before
coming to a caustic. Then \eq{boussobound} is modified
as\cite{wald3}
\bean S_L\leq\frac{1}{4}|\abp-A_{B}|\,.
\label{gbb} \eean This bound is obviously stronger than the original
Bousso bound and its validity has been proved by assuming some
physical conditions.

\begin{figure}[htmb]
\centering \scalebox{0.7} {\includegraphics{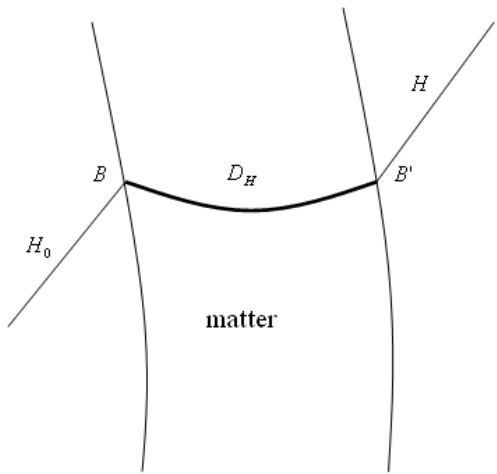}} \caption{When
matter collapses to an existing black hole, the dynamical horizon
$D_H$ is formed, which connects the old horizon $H_0$ and the new
horizon $H$. } \label{kruskal}
\end{figure}

Recently, an entropy bound related to a dynamical horizon has been
discussed \cite{zhang1,zhang2}. A dynamical horizon is a spacelike
hypersurface foliated by closed, marginally trapped 2-spheres, which
describes an intermediate state when a black hole is
formed\cite{ashtekar}. Suppose that some matter is collapsing into
an existing black hole. In Fig. 1, $H_0$ represents the horizon of
the existing black hole.  A dynamical horizon $D_H$ begins to form
at moment $B$. At moment $B'$, the dynamical horizon finally evolves
to an event horizon $H$ and a new black hole is formed. Since each
point in Fig. 1 represents a closed 2-surface, we denote the areas
at $B$ and $B'$ by $A_B$ and $A_B'$ respectively. Let $S_D$ be the
entropy flux through the dynamical horizon. It has been shown in
\cite{ashtekar} that the area of the two-surface increases
monotonically along $D_H$. It then follows that $\abp>A_B$, meaning
that entropy of the black hole must increase and the second law of
black hole mechanics holds \cite{ashtekar}. However, this result is
not enough to guarantee the generalized second law since the entropy
of the matter has not been counted. Because the matter losses all
its entropy after the formation of the new horizon $H$, the
generalized second law is equivalent to the
following bound 
\bean S_D\leq \frac{1}{4}(\abp-A_B)\,. \label{dbound} \eean This
bound was given as a conjecture in \cite{zhang2} and its validity
was tested only by a specific example. The purpose of our paper is
to give a general proof for the conjecture. Apparently, the bounds
\meq{gbb} and \meq{dbound} have the similar form. However, they are
different in nature because the entropy in \eq{gbb} crosses a null
surface and the entropy in \eq{dbound} is contained in a spacelike
dynamical horizon. We have mentioned that the spacelike entropy
bounds \meq{bbound} and \meq{areabound} can be easily violated for
sufficiently large volumes where self-gravitation is strong. As a
spacelike entropy bound, the validity of \eq{dbound} is not obvious.
On the other hand, the Bousso bound \meq{boussobound} and its
generalized version \meq{gbb} has passed all the tests so far in
classical regimes. After investigating the relations between the
generalized Bousso bound \meq{gbb} and the spacelike bound
\meq{dbound}, we find that the bound \meq{dbound} is actually
implied by the well-tested bound \meq{gbb}. Our proof in the next
section will be based on the generalized Bousso bound as well as the
properties of the dynamical horizon.

\section{Proof of the entropy bound related to a dynamical
horizon} \label{proof}
In order to prove the bound \meq{dbound} in a
model as shown in
Fig. 1, we make the following assumptions: \\
 1. The generalized Bousso bound \meq{gbb} holds. \\
 2. The ordinary second law holds for the matter that crosses the horizon.\\
 3. The spacetime is spherically symmetric.\\
 4. The dynamical horizon is a future outer trapped horizon (FOTH)
 defined by Hayward\cite{hayward}.

 The first assumption is essential in the following proof. The
 second assumption is quite natural. Assumption 3 simplifies the
 structure of the spacetime. The relevant dimension of spacetime is then reduced to
 two when only radial null geodesics are considered.
 We shall show later that this  restriction can be released under certain
 conditions. We shall not use the original definition of dynamical
 horizon proposed by Ashtekar {\em et.al.} in \cite{ashtekar}.
 Instead, we use a close but more restricted definition as stated in
 assumption 4. As proposed by Hayward\cite{hayward}, a FOTH is a
 spacelike hypersurface which is foliated by a family of closed
 two-surfaces, such that on each leaf \\
 (i) the expansion of one future directed null normal $l^a$
 vanishes, i.e., $\theta_l=0$. \\
 (ii)the expansion of the other future directed null normal $n^a$ is
 negative, $\theta_n=0$.\\
 (iii) ${\cal L}_n \theta_{l}<0$\\

The condition (iii) makes the FOTH differ from Ashtekar's dynamical
horizon because it provides more information in a neighborhood of
the horizon. As proven by Hayward, any spatial two-surface
sufficiently close to a FOTH and in its past is not trapped. This
property will be used in our following proof.

\begin{figure}[htmb]
\centering \scalebox{0.7} {\includegraphics{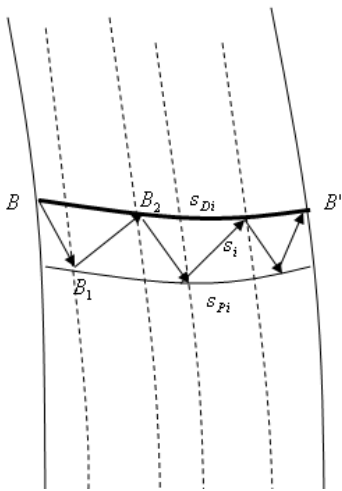}}
\caption{The short lines with arrows represent the ingoing and
outgoing null geodesics. The dotted lines represent the comoving
observers.}
\label{kruskal}
\end{figure}
Since we are considering a dynamical horizon (FOTH) $D_H$, it
follows by definition that $D_H$ is foliated by a family of closed
2-surfaces. Now consider a spacelike hypersurface, $D_P$, in the
past of $D_H$ (see Fig. 2). A null geodesic starting from $B$
propagates backward to the past and bounces off at $B_1$ on $D_P$.
If the light continues to propagate in this way, it will finally
reach $B'$ (or some point on $D_H$ close to $B'$). We denote each
point (representing a 2-surface) by $B_i$ and its area by $A_i$.
Since each section of the null line, denoted by $h_i$, represents a
null hypersurface connecting two 2-surfaces, we may apply the
generalized Bousso bound to each one of them. Note that the
dynamical horizon is essentially the boundary of the trapped region
and $D_P$ lies in its past. Therefore, according to Hayward's
theorem, the foliations on $D_P$ are not trapped if $D_P$ is
sufficiently close to $D_H$, which means that the expansion
$\theta_i$ of each null generators of $h_i$ has a unique sign. For
example, the null generators $B_1\rightarrow B$ and $B_1\rightarrow
B_2$ have negative and positive expansions, respectively.
Consequently, we have $A_B<A_{B_1}<A_{B_2}...<A_{B'}$. Denote the
entropy crossing each small null surface by $s_i$. Then the bound
\meq{gbb} yields
\bean
s_1&\leq& \frac{1}{4}(A_{B_1}-A_{B}) \nonumber \\
s_2&\leq& \frac{1}{4}(A_{B_2}-A_{B_1}) \nonumber \\
&......&  \nonumber\\
s_n&\leq& \frac{1}{4}(A_{B'}-A_{B_{n-1}}) \label{agbb} \eean By
adding all the inequalities above, we find
\bean \sum_{i=1}^n s_i\leq\frac{1}{4}(A_{B'}-A_B) \,.\label{sin}
\eean Now we have obtained the right-hand side of the bound
\meq{dbound}. But the left-hand side of \meq{sin} is the total
entropy crossing all the small null surfaces, not the entropy
crossing the dynamical horizon. To proceed, consider comoving
observers passing through the connecting surfaces of the null
geodesics (the dotted lines in Fig.2.). Denote the entropy that
flows through an interval between two adjacent observers on $D_P$ by
$s_{Pi}$ and the corresponding one on $D_H$ by $s_{Di}$. By using
our assumption 2, the ordinary second law, we have immediately
\bean s_{Pi}\leq s_i\leq s_{Di}
\label{three} \eean and by summation, we obtain 
\bean S_P\leq \sum_{i=1}^n s_i\ \leq S_D \,, \label{sum} \eean where
$S_P$ is the entropy crossing $D_P$. To complete the proof, we need
to replace the left-hand side in \eq{sin} by $S_D$. Obviously,
inequality \meq{sum} does not give the desired direction. This can
be remedied by the following observation. Since our argument applies
to any spacelike surface $D_P$ in the past of $D_H$, it applies to
$D_P$ that approaches to $D_H$. When $D_P$ gets arbitrarily close to
$D_H$, we have $S_P\sim S_D$. By virtue of \eq{sum}, we have $\sum
s_i\ \sim S_D$. Note that the right-hand side of \eq{sin} is a fixed
value, which is independent of the limiting process. On the other
hand, $S_D$ is a limit of $S_P$ and consequently a limit of $\sum
s_i$. Therefore, if \eq{sin} holds for $\sum s_i$, it  certainly
holds for its limit $S_D$, as we desired to show.

\section{Gerneralizations of the proof}
Now we shall generalize our proof above from spherically symmetric
case to non-spherically symmetric case and from a dynamical horizon
to a general spacelike hypersurface.
\begin{figure}[htmb]
\centering \scalebox{0.7} {\includegraphics{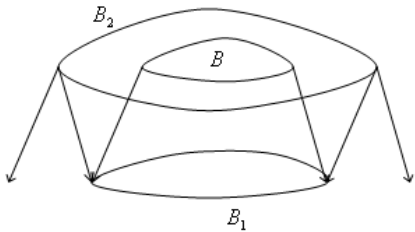}} \caption{$B$
and $B_2$ are closed two-surfaces on $D_H$. $B_1$ is the
intersection of the two light-sheets starting from $B$ and $B_2$. }
\label{non-sphere}
\end{figure}

\subsection{non-spherically symmetric case}
In the proof above, we see that spherical symmetry plays an
important role. It guarantees that the light rays starting from $B$
will intersect $D_P$ and $D_H$ as two-spheres (i.e., $B_1$, $B_2$...
in Fig.2 are all two-spheres). Therefore, when $D_P$ is chosen close
enough to $D_H$, the final cross-section of the light rays will
always coincide with $B'$ as shown in Fig.2. If the spherical
symmetry is not imposed, the shape of the cross-sections will be out
of control and the final cross-section will not coincide with $B'$
in general. However, we can fix the problem by the following
procedure. In the previous proof, the foliations on $D_H$ are the
intersections of the light-sheets and $D_H$. In non-spherical cases,
we may fix the foliations first, i.e., choosing $B$, $B_2$, ...$B'$,
on $D_H$. Then let the light rays propagate back to the past from
each of these foliations inwardly and outwardly (see
Fig\ref{non-sphere}). We assume that the intersections of the
three-dimensional lightsheets are 2-dimensional spacelike surfaces,
labeled by $B_1$, $B_3$,...Then our proof for spherical cases can be
applied similarly. Therefore, the entropy bound \meq{dbound} is
generalized to non-spherically symmetrically dynamical horizons.
Note that we have made additional assumption in the generalization,
requiring the intersection of two lightsheets be a spacelike
two-surface. This should not be a strong condition  since we may
choose the spacelike foliations on $D_H$ arbitrarily close.

\subsection{The bound on a spacelike hypersurface}
Our discussion above has been aimed at dynamical horizons. However,
the relevant property we used is that there is a non-trapped region
in a neighborhood of $D_H$ such that the inequalities \meq{agbb}
hold. It is not difficult to see that our proof applies to any
spacelike hypersurface which lies outside a trapped or anti-trapped
region. Thus, the lightlike Bousso bound \meq{gbb} implies the
spacelike entropy bound \meq{dbound}. Indeed, we may treat them as
one unified entropy bound, which holds for lightlike hypersurface
with negative expansion and spacelike hypersurface where no trapped
or anti-trapped two-surface exists.

\section{Conclusions}

The entropy bound conjecture discussed in \cite{zhang1} and
\cite{zhang2} has been proved based on the generalized Bousso bound
and the ordinary second law. Therefore,  the generalized second law
is satisfied in the presence of a dynamical horizon. We also showed
that the lightlike entropy bound and the spacelike entropy bound can
be unified as one bound.

\section*{Acknowledgements}
S.G. was supported in part by NSFC Grants 10605006, and the
Scientific Research Foundation for the Returned Overseas Chinese
Scholars, State Education Ministry. X.Wu was supported in part by
NSFC Grants 10705048, 10605006, 10731080.

\end{document}